\documentclass{article}
\usepackage[utf8]{inputenc} 
\usepackage[T1]{fontenc}
\usepackage{spconf,amsmath,graphicx}
\usepackage{enumitem}
\setlist{nosep, leftmargin=14pt}
\usepackage{mwe} 
\usepackage{hyperref}       
\usepackage{url}            
\usepackage{booktabs}       
\usepackage{amsfonts}       
\usepackage{nicefrac}       
\usepackage{microtype}      
\usepackage{lipsum}
\usepackage{stackengine}
\usepackage{float}
\usepackage{amssymb}
\usepackage{subfig}
\usepackage{float}
\usepackage{color, colortbl}
\usepackage{rotating}
\usepackage{subfig}
\usepackage{tikz}
\usepackage{xcolor}
\usepackage{animate}
\usepackage{mathrsfs}

\usepackage[ruled,vlined]{algorithm2e}

\hypersetup{
    colorlinks,
    linkcolor={red!50!black},
    citecolor={blue!80!black},
    urlcolor={blue!80!black}
}

\title{Correlation of Correlation Networks: High-Order Interactions in the Topology of Brain Networks}
\name{Qiang Li\thanks{qli27@gsu.edu \\ \textbf{\textit{IEEE International Symposium on Biomedical Imaging (ISBI 2025)}}}, Jingyu Liu, Vince D. Calhoun}
\address{Tri-Institutional Center for Translational Research in Neuroimaging and Data Science (TReNDS), \\
Georgia State, Georgia Tech, and Emory University, Atlanta, GA 30303, United States}

\begin{document}
\maketitle

\begin{abstract}
To understand collective network behavior in the complex human brain, pairwise correlation networks alone are insufficient for capturing the high-order interactions that extend beyond pairwise interactions and play a crucial role in brain network dynamics. These interactions often reveal intricate relationships among multiple brain networks, significantly influencing cognitive processes. In this study, we explored the correlation of correlation networks and topological network analysis with resting-state fMRI to gain deeper insights into these higher-order interactions and their impact on the topology of brain networks, ultimately enhancing our understanding of brain function. We observed that the correlation of correlation networks highlighted network connections while preserving the topological structure of correlation networks. Our findings suggested that the correlation of correlation networks surpassed traditional correlation networks, showcasing considerable potential for applications in various areas of network science. Moreover, after applying topological network analysis to the correlation of correlation networks, we observed that some high-order interaction hubs predominantly occurred in primary and high-level cognitive areas, such as the visual and fronto-parietal regions. These high-order hubs played a crucial role in information integration within the human brain.
\end{abstract}


\section{Introduction}
\label{sec:intro}
Correlation networks are commonly used to measure pairwise statistical relationships in network science, often referred to as \textit{functional connectivity} in neuroscience when applied to functional MRI time series ~\cite{Yeo11,Karl11Brainc}. They primarily explain the interactions between pairwise brain regions. However, when it comes to more complicated features embedded in brain signals, such as statistically higher order interactions or nonlinear relations, correlation networks frequently miss essential information~\cite{Battiston20,Li22,QiangEntr22,QiangNC23}. High-order interactions, which go beyond pairwise correlations, can reveal detailed connectivity patterns that pairwise correlations cannot capture~\cite{Li22, QiangEntr22, QiangNC23}. This has a significant impact on our understanding of brain dynamics. 

Recognizing the significance of high-order interactions, researchers have gradually shifted more attention to this area~\cite{Battiston20,Li22,QiangEntr22}. Several metrics have already been employed in brain network studies, yielding intriguing results that pairwise metrics typically do not capture. 

From an \textit{information-theoretical perspective}, total correlation~\cite{Watanabe60} and dual total correlation~\cite{Han78} are two metrics that can be employed to estimate interactions beyond pairwise relationships. Both metrics have been used to explore high-order interactions in healthy brains~\cite{Li22,QiangNC23} and in various brain diseases~\cite{QiangEntr22,Herzog22}. They capture new connections that are often not identified through pairwise interactions.

From a \textit{graph theory perspective}, hypergraphs represent a significant approach to understanding high-order interactions within graphs. When combined with graph neural networks, they have increasingly been applied in studies related to brain network science~\cite{Xiao2019tmi}.

From the \textit{topology and geometric data analysis perspective}, techniques such as persistent homology, Euler characteristic, and curvature have been employed to quantify high-order interactions~\cite{SantosPhysRevE19,Zampieri22Bsf}. These methods have indeed yielded new insights that are often overlooked in pairwise analyses. 

In this study, we explored correlation of correlation networks, which are derived from correlation networks, and combined topological network analysis techniques to better capture high-order networks in the brain. Our findings indicate that correlation of correlation networks outperforms traditional correlation networks, demonstrating significant potential for application in other fields of network science. Moreover, applying topological network analysis to the correlation of correlation networks will further enhance our ability to identify high-order hubs in the brain.

\section{Methodology}
\label{sec:meths}
\subsection{1000 Functional Connectomes Dataset}
The 1000 Functional Connectomes Project is a comprehensive collection of resting state fMRI datasets from over 1000 participants, gathered across more than 30 independent studies worldwide~\cite{Biswal10pnas,Kalcher12finhn}. The final sample included 806 subjects (see Tab.\ref{demog}), randomly selected from the remaining dataset, which comprises 33 independent samples from 26 research centers located in North America (15), Europe (8), Asia (2), and Australia (1). 

\begin{table}[!htbp]
\centering
\caption{\textcolor{black}{Demographic Statistics for Selected Samples.}}
\label{demog}
\resizebox{\columnwidth}{!}{%
\begin{tabular}{llllll}
\hline
\multicolumn{1}{l|}{\textbf{Sites}} & \multicolumn{1}{l|}{\textbf{Subjects}} & \multicolumn{1}{l|}{\textbf{Mean Age $\pm$ SD Age}} & \multicolumn{1}{l|}{\textbf{Voxel Size}} & \multicolumn{1}{l|}{\textbf{TR}} & \textbf{Volumes} \\ \hline
Baltimore & 23 & 29.3 $\pm$ 5.5 & 21.3 & 2.50 & 123 \\ \hline
Bangor & 20 & 23.4 $\pm$ 5.3 & 27.0 & 2.00 & 265 \\ \hline
Berlin & 26 & 29.8 $\pm$ 5.2 & 36.0 & 2.30 & 195 \\ \hline
Cambridge & 198 & 21.0 $\pm$ 2.3 & 27.0 & 3.00 & 119 \\ \hline
Cleveland & 31 & 43.5 $\pm$ 11.1 & 16.0 & 2.80 & 127 \\ \hline
Dallas & 24 & 42.6 $\pm$ 20.1 & 47.3 & 2.00 & 115 \\ \hline
ICBM & 86 & 44.2 $\pm$ 17.9 & 27.0 & 2.00 & 192 \\ \hline
Leiden\_2180 & 12 & 23.0 $\pm$ 2.5 & 40.7 & 2.18 & 215 \\ \hline
Leipzig & 37 & 26.2 $\pm$ 5.0 & 36.0 & 2.30 & 195 \\ \hline
Milwaukee\_a & 18 & NA & 84.4 & 2.00 & 175 \\ \hline
Munchen & 16 & 68.4 $\pm$ 4.0 & 43.0 & 3.00 & 72 \\ \hline
Newark & 19 & 24.1 $\pm$ 3.9 & 59.1 & 2.00 & 135 \\ \hline
NewYork\_a & 84 & 35.0 $\pm$ 9.6 & 27.0 & 2.00 & 192 \\ \hline
Orangeburg & 20 & 40.6 $\pm$ 11.0 & 61.2 & 2.00 & 165 \\ \hline
Oulu & 103 & 21.5 $\pm$ 0.6 & 70.4 & 1.80 & 245 \\ \hline
Oxford & 22 & 29.0 $\pm$ 3.8 & 31.5 & 2.00 & 175 \\ \hline
PaloAlto & 17 & 32.5 $\pm$ 8.1 & 57.9 & 2.00 & 235 \\ \hline
Queensland & 19 & 25.9 $\pm$ 3.9 & 46.5 & 2.10 & 190 \\ \hline
SaintLouis & 31 & 25.1 $\pm$ 2.3 & 64.0 & 2.50 & 127 \\ \hline
\end{tabular}%
}
\end{table}

The original scans were conducted using echo planar imaging (EPI) during resting-state sessions, utilizing variable scanning parameters.

\subsection{Correlation of Correlation Networks}
After standard preprocessing of the resting state fMRI (rsfMRI), time courses were extracted based on 177 sets of identified coordinates. For an fMRI signal of \(\mathscr{N}\) regions, where \(x(i)\) (\(1 \leq i \leq \mathscr{N}\), \(\mathscr{N} = 177\)), we can informally represent this as \(\mathscr{N} = (x_{1}(t), x_{2}(t), x_{3}(t), \ldots, x_{\mathscr{N}}(t))\) points with a constant time interval \(t\). A correlation matrix of Pearson values was derived from the time series data, and Fisher's z transformation was applied to the correlation coefficients of each of the 177 brain regions with every other region. Consequently, the correlation networks, as measured using traditional Pearson correlation, are presented as follows: 

\begin{equation}
\nonumber
    CN\rho_{i j}=\frac{\sum_{t=1}^T\left(x_i(t)-\hat{x_i}\right)\left(x_j(t)-\hat{x_j}\right)}{\sqrt{\sum_{t=1}^T\left(x_i(t)-\hat{x_i}\right)^2 \sum_{t=1}^T\left(x_j(t)-\hat{x_j}\right)^2}}
\end{equation}
where $\hat{x_i}$ and $\hat{x_j}$ are the average of the rsfMRI signals at regions $i$ and $j$. 

The correlation networks from the selected sample are available at the UCLA Multimodal Connectivity Database (\url{http://umcd.humanconnectomeproject.org}).

Then, the correlation of correlation networks can be derived from the correlation networks~\cite{Zhang2017TestRetestRO,Qiang24CCN}. Letting \(CN\rho_{i}\) represent the correlation network profile for region \(i\), we define \(CN\rho_{i\cdot} = \left\{CN\rho_{ik \mid k \in \mathbb{R}_{1}^{177}, k \neq i}\right\}\). Thus, the correlation of correlation networks can be denoted as follows:

\begin{equation}
\nonumber
   CCN_{ij}=\frac{\sum_k\left(CN\rho_{i k}-CN\hat{\rho}_i \cdot\right)\left(CN\rho_{j k}-CN\hat{\rho_j} \cdot\right)}{\sqrt{\sum_k\left(CN\rho_{i k}-CN\hat{\rho_i} \cdot\right)^2} \sqrt{\sum_k\left(CN\rho_{j k}-CN\hat{\rho_j} \cdot\right)^2}}
\end{equation}

Here, the 177 brain regions can be categorized into six intrinsic functional subnetworks of the human brain: the Default Mode Network (DMN), Visual Network (VIS), Fronto-Parietal Task Control Network (FP), Ventral Attention Network (VA), Sensorimotor Network (SM), and Limbic Network (LIM).

\subsection{Topological Data Analysis for Correlation of Correlation Networks}

Topological data analysis (TDA) captures various network characteristics by examining the high-order structures of a network, going beyond the pairwise connections typically used in graph theory~\cite{Zampieri22Bsf}. Additionally, it addresses the thresholding problem found in correlation networks. This investigative process, known as filtration, explores all possible thresholds rather than selecting a fixed one~\cite{Zampieri22Bsf}. It involves adjusting the density \(d\) of the network, where \(0<d<1\). This results in a nested sequence of networks; as \(d\) increases, the network becomes more densely connected. Through filtration, information about high-order interactions in complex brain networks can be revealed.

\section{Results}
\label{sec:res}
\begin{figure*}[!htbp]
    \centering
    \includegraphics[width=0.88\textwidth, height=10.6cm]{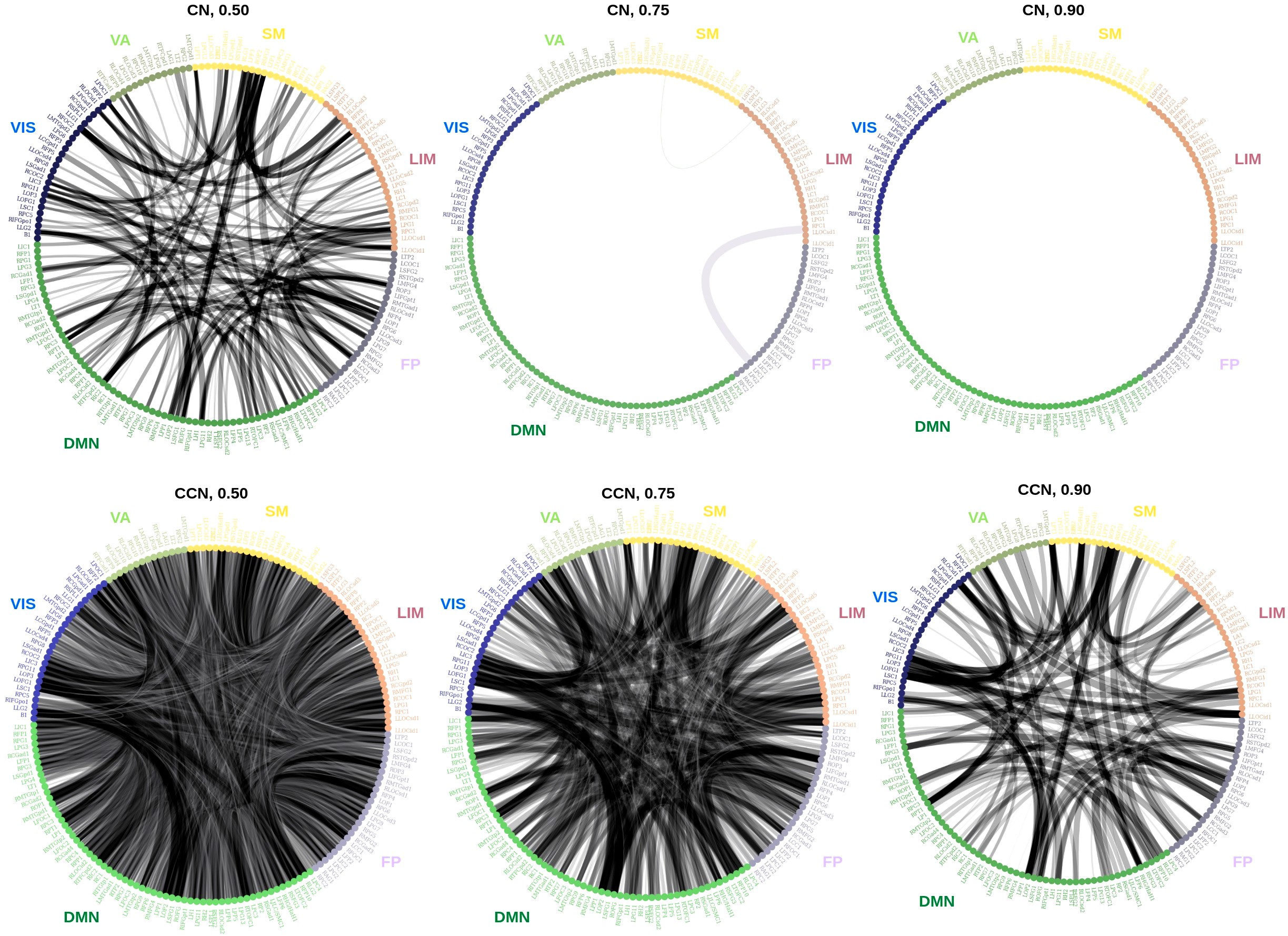}
    \caption{\textbf{Correlation networks (CN) vs. correlation of correlation networks (CCN).} The first row displays correlation networks at various thresholds (0.50, 0.75, and 0.90). The second row illustrates the correlation of correlation networks using the same thresholds. The 177 brain regions are categorized into six sub-networks (DMN, FP, LIM, SM, VA, and VIS).}
    \label{fig:1}
\end{figure*}

In Fig.\ref{fig:1}, the brain connectome is plotted using correlation networks and correlation of correlation networks at different thresholds (0.50, 0.75, and 0.90). Compared to correlation networks, we clearly observe that correlation of correlation networks reveals stronger connections and highlights weaker connections that may be overlooked. For example, at thresholds of 0.75 and 0.90, most connection information disappears in correlation networks, whereas correlation of correlation networks retains much of this information. 

\begin{figure*}[!htbp]
    \centering
    \includegraphics[width=0.8\textwidth, height=8.7cm]{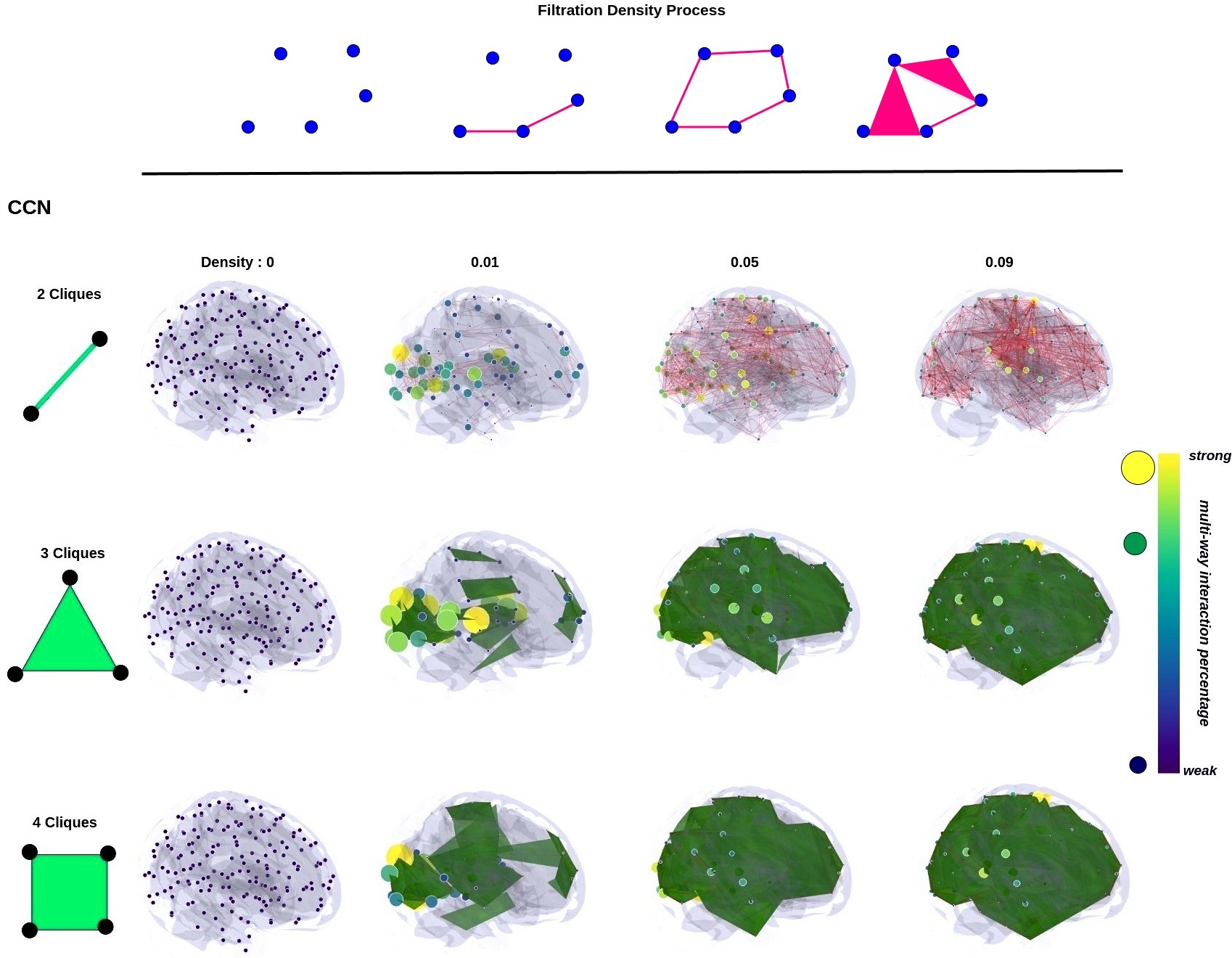}
    \caption{\textbf{The correlation of correlation networks (CCN) in topology space.} Here, we illustrate high-order interactions involving 2, 3, and 4 cliques, demonstrating how different orders are formed in the brain throughout the filtration density process (from left to right: 0, 0.01, 0.05, and 0.09). The size of the nodes reflects the frequency of involvement in higher-order interactions.}
    \label{fig:2}
\end{figure*}

To further explore high-order interactions in the human brain, we applied a filtration process for different orders of interactions (2, 3, and 4), noting that higher orders consume more time and computational power. As the density increases from 0 to 0.09, we observe a greater density of connections in the human brain, as illustrated in Fig.\ref{fig:2}. Moreover, we observe that some high-order interaction hubs occur predominantly in primary and high-level cognitive areas, such as the visual and fronto-parietal regions. These hubs may play a crucial role in complex cognitive processes, integrating information from various sensory modalities and facilitating higher-order functions. Their prominence in these regions underscores the importance of network interactions in supporting advanced cognitive capabilities in the human brain.

\section{Conclusion and Discussion}
The results suggest that correlation of correlation networks indeed provides more connection information compared to standard correlation networks. This enhancement allows us to capture more connection features, which is beneficial for classification in deep learning, as it offers richer feature information. Meanwhile, we applied topological data processing to the correlation of correlation networks, capturing high-order hubs primarily in the visual and fronto-parietal regions. In summary, the correlation of correlation networks combined with topological network analysis will provide a new way to exploring high-order information interactions in human brain.

However, we recognize that brain activity is not static. As a result, the correlation of correlation networks has limitations in capturing dynamic high-order correlation networks. Furthermore, additional high-order network statistical properties should be measured in the future.

\section{Acknowledgements}
The authors declare that there are no conflicts of interest related to this research. This work was supported by NSF grant 2112455, and NIH grants R01MH123610 and R01MH119251.
\bibliographystyle{ieeetr}
\bibliography{references} 

\begin{thebibliography}{10}

\bibitem{Yeo11}
B.~T. Thomas~Yeo, F.~M. Krienen, J.~Sepulcre, M.~R. Sabuncu, D.~Lashkari, M.~Hollinshead, J.~L. Roffman, J.~W. Smoller, L.~Zöllei, J.~R. Polimeni, B.~Fischl, H.~Liu, and R.~L. Buckner, ``The organization of the human cerebral cortex estimated by intrinsic functional connectivity,'' {\em Journal of Neurophysiology}, vol.~106, no.~3, pp.~1125--1165, 2011.

\bibitem{Karl11Brainc}
K.~Friston, ``Functional and effective connectivity: A review,'' {\em Brain connectivity}, vol.~1, pp.~13--36, 01 2011.

\bibitem{Battiston20}
F.~Battiston {\em et~al.}, ``Networks beyond pairwise interactions: Structure and dynamics,'' {\em Physics Reports}, vol.~874, pp.~1--92, 2020.

\bibitem{Li22}
Q.~Li, ``Functional connectivity inference from fmri data using multivariate information measures,'' {\em Neural Networks}, vol.~146, pp.~85--97, 2022.

\bibitem{QiangEntr22}
Q.~Li, G.~V. Steeg, S.~Yu, and J.~Malo, ``Functional connectome of the human brain with total correlation,'' {\em Entropy}, vol.~24, no.~12, 2022.

\bibitem{QiangNC23}
Q.~Li, G.~Ver~Steeg, and J.~Malo, ``Functional connectivity via total correlation: Analytical results in visual areas,'' {\em Neurocomputing}, vol.~571, p.~127143, 12 2023.

\bibitem{Watanabe60}
S.~Watanabe, ``Information theoretical analysis of multivariate correlation,'' {\em IBM Journal of research and development}, vol.~4, no.~1, pp.~66--82, 1960.

\bibitem{Han78}
T.~S. Han, ``Nonnegative entropy measures of multivariate symmetric correlations,'' {\em Inf. Control.}, vol.~36, no.~2, pp.~133--156, 1978.

\bibitem{Herzog22}
R.~Herzog, F.~Rosas, R.~Whelan, S.~Fittipaldi, H.~Santamaria-Garcia, J.~Cruzat, A.~Birba, S.~Moguilner, E.~Tagliazucchi, P.~Prado, and A.~Ibanez, ``Genuine high-order interactions in brain networks and neurodegeneration,'' {\em Neurobiology of Disease}, vol.~175, p.~105918, 2022.

\bibitem{Xiao2019tmi}
L.~Xiao, J.~Wang, P.~H. Kassani, Y.~Zhang, Y.~Bai, J.~M. Stephen, T.~W. Wilson, V.~D. Calhoun, and Y.~ping Wang, ``Multi-hypergraph learning-based brain functional connectivity analysis in fmri data,'' {\em IEEE Transactions on Medical Imaging}, vol.~39, pp.~1746--1758, 2019.

\bibitem{SantosPhysRevE19}
F.~A.~N. Santos, E.~P. Raposo, M.~D. Coutinho-Filho, M.~Copelli, C.~J. Stam, and L.~Douw, ``Topological phase transitions in functional brain networks,'' {\em Phys. Rev. E}, vol.~100, p.~032414, Sep 2019.

\bibitem{Zampieri22Bsf}
E.~Zampieri, G.~Moreni, C.~Vriend, L.~Douw, and F.~Santos, ``A hands-on tutorial on network and topological neuroscience,'' {\em Brain Structure and Function}, vol.~227, 04 2022.

\bibitem{Biswal10pnas}
B.~Biswal, M.~Mennes, X.-N. Zuo, S.~Gohel, C.~Kelly, S.~Smith, C.~Beckmann, J.~Adelstein, R.~Buckner, S.~Colcombe, A.-M. Dogonowski, M.~Ernst, D.~Fair, M.~Hampson, M.~Hoptman, J.~Hyde, V.~Kiviniemi, R.~Kötter, S.-J. Li, and M.~Milham, ``Toward discovery science of human brain function,'' {\em Proceedings of the National Academy of Sciences of the United States of America}, vol.~107, pp.~4734--9, 03 2010.

\bibitem{Kalcher12finhn}
K.~Kalcher, W.~Huf, R.~Boubela, P.~Filzmoser, L.~Pezawas, B.~Biswal, S.~Kasper, E.~Moser, and C.~Windischberger, ``Fully exploratory network independent component analysis of the 1000 functional connectomes database,'' {\em Frontiers in Human Neuroscience}, vol.~6, p.~301, 11 2012.

\bibitem{Zhang2017TestRetestRO}
H.~Zhang, X.~Chen, Y.~Zhang, and D.~Shen, ``Test-retest reliability of “high-order” functional connectivity in young healthy adults,'' {\em Frontiers in Neuroscience}, vol.~11, 2017.

\bibitem{Qiang24CCN}
Q.~Li, W.~Huang, C.~Qiao, and H.~Chen, ``Unraveling integration-segregation imbalances in schizophrenia through topological high-order functional connectivity,'' {\em BioRxiv:10.1101/2024.10.03.616506}, 10 2024.

\end{thebibliography}
\end{document}